\documentclass[letterpaper, 10 pt, conference]{ieeeconf}  

\IEEEoverridecommandlockouts                              
\title{\LARGE \bf
Decoding Emotional Experience through \\
Physiological Signal Processing}

\author{Maria S. Perez-Rosero$^{1}$, Behnaz Rezaei$^{1}$, Murat Akcakaya$^{2}$, and Sarah Ostadabbas$^{1}$ 
\thanks{$^{1}$Maria S. Perez-Rosero, Behnaz Rezaei, and Sarah Ostadabbas are with the Augmented Cognition laboratory, Electrical and Computer Engineering Department, Northeastern University, MA, USA (corresponding author's e-mail: ostadabbas@ece.neu.edu).}
\thanks{$^{2}$Murat Akcakaya is with the Electrical and Computer Engineering Department, University of Pittsburgh 
PA, USA.}
}

\usepackage{array}
\usepackage{ifpdf}
\usepackage{graphicx}
\usepackage{epsfig} 

\usepackage[cmex10]{amsmath}
\usepackage{amsfonts}
\usepackage{algorithmic}
\usepackage{algorithm2e}
\usepackage{mdwmath}
\usepackage{colortbl}
\usepackage{hhline}
\usepackage{multirow}
\usepackage[caption=false,font=footnotesize]{subfig}
\usepackage{url}
\usepackage{fixltx2e}
\usepackage{float}
\usepackage{balance}
\usepackage[english]{babel}
\usepackage{comment}

\newcommand{\figref}[1]{Fig.~\ref{fig:#1}}
\newcommand{\tblref}[1]{Table~\ref{tbl:#1}}

\newcommand{\figsignals}{
\begin{figure}
  \centering
  \includegraphics[width=\linewidth,trim=0in 0in 0in 0in,
  clip=true]{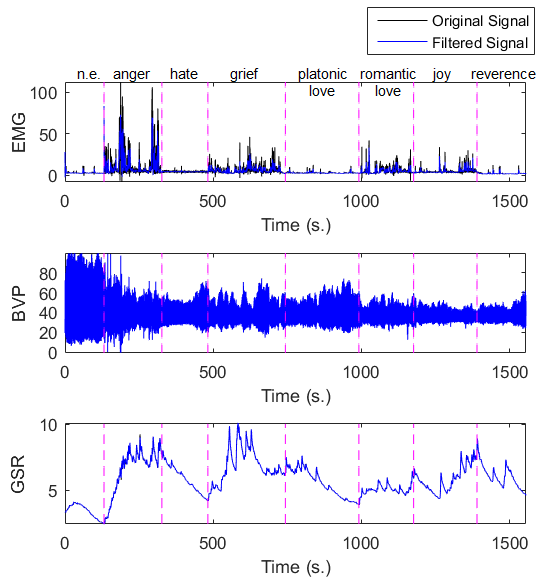}
  \caption{From top to bottom, examples of the three physiological signals under study obtained from dataset II in \cite{ref1}: \textit{electromyogram} (microvolts), \textit{blood volume pressure} (percent reflectance), \textit{galvanic skin response} (micro Siemens). These signals were measured while the participant experienced the eight emotions proposed for recognition. The segments of data corresponding to each emotion are divided by the magenta dashed lines and its corresponding labels are placed at the top of the graph.}
\label{fig:signals}
\end{figure}
}

\newcommand{\figweakAlg}{
\begin{figure*}
  \centering
  \includegraphics[width=0.96\textwidth,trim=0in 0in 0in 0in,
  clip=true]{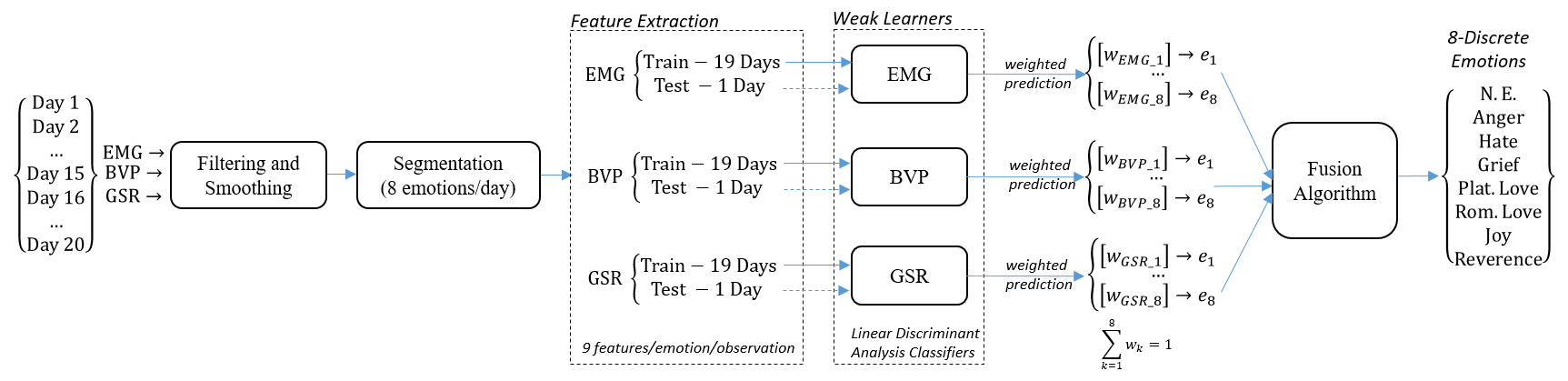}
  \caption{Emotion classification algorithm based on fusion of weak learners. Twenty-day observations of three physiological signals reported in \cite{ref1} are the input to a pre-processing block constituted of filtering, smoothing and segmentation procedures. These pre-processed signals are divided into training set (19-day observations) and test set (1-day observation). For each of these sets, nine features were computed per emotion per observation day. Three specialized weak learners based on linear discriminant classifiers are build for each physiological signal. As a final step, the weighted predictions provided as the outputs of each weak learner are fused to decode eight discrete emotions.}
\label{fig:weakAlg}
\end{figure*}
}

\newcommand{\figfusionAlg}{
\begin{figure*}
  \centering
  \includegraphics[width=0.96\textwidth,trim=0in 0in 0in 0in,
  clip=true]{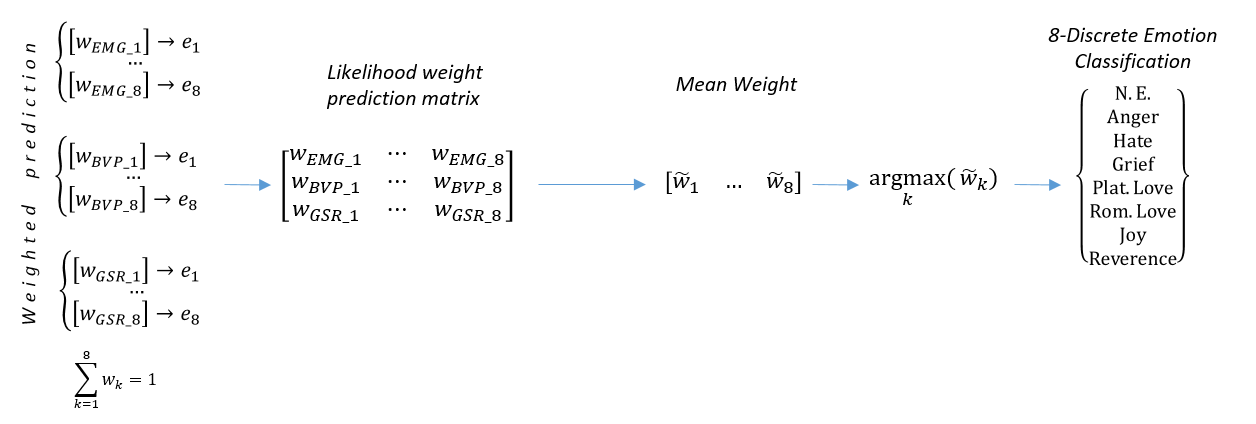}
  \caption{Fusion Algorithm -- Given a test input containing the EMG, BVP and GSR signals, each weak learner provides a weighted prediction of the 8 emotions that the input is likely to be constituted of. Next, all the weighted predictions from the 3 weak learners are concatenated in a matrix in order to compute a mean weigh vector that will consolidate these three different criteria. Finally, an discrete 8-emotion classification is obtained by computing the emotion for which the maximum mean weight was computed.}
\label{fig:fusionAlg}
\end{figure*}
}

\newcommand{\figtprFpr}{
\begin{figure}
  \centering
  \includegraphics[width=0.5\textwidth,trim=0in 0in 0in 0in,
  clip=true]{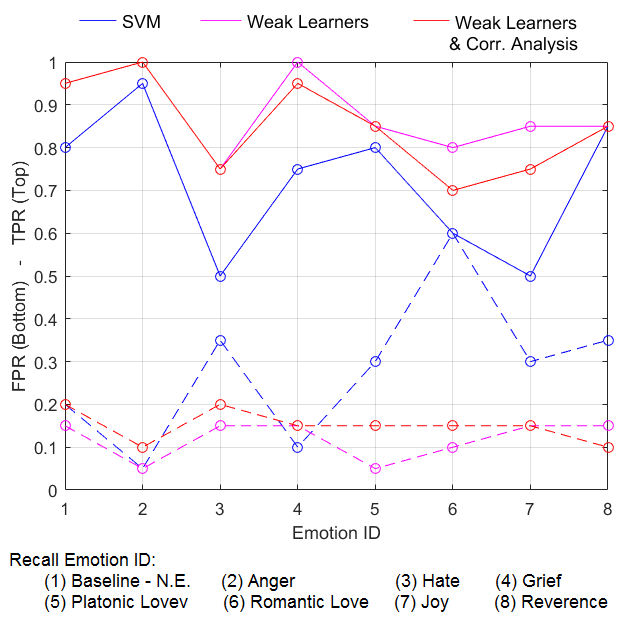}
  \caption{
    True Positive Rate -- TPR (solid lines) and False Positive Rate -- FPR (dashed lines) curves for each of the eight target emotions under the scope of the three classifiers under comparison. Note that the proposed weak learners with and without correlation analysis outperform SVM by achieving a higher TPR and lower FPR.}
\label{fig:tprFpr}
\end{figure}
}

\newcommand{\figmisclassification}{
\begin{figure*}
  \centering
  \includegraphics[width=1.0\textwidth,trim=0in 0in 0in 0in,
  clip=true]{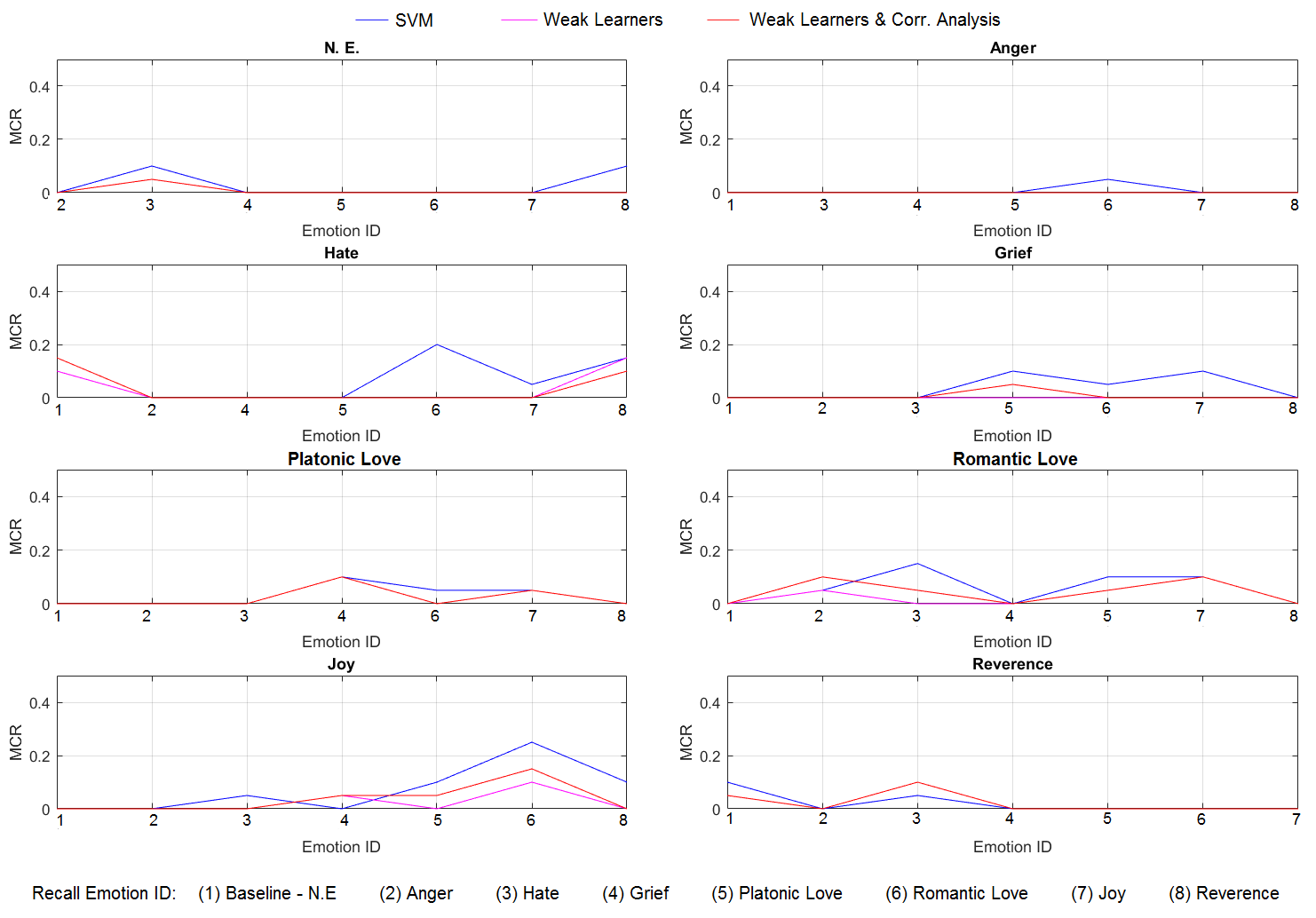}
  \caption{Misclassification rates (MSCR) for each of the eight target emotions under analysis when compared with the other seven remaining emotions. For instance, it can be observed that the emotion with the lowest TPR of 0.7, romantic love, is misclassified mainly with anger, joy and in lesser degree by hate and platonic love.}
\label{fig:misclassification}
\end{figure*}
}

\newcommand{\tblaccuracy}{
\begin{table}[b]
  \centering
  \caption{Comparison between different classifiers' recognition performance} 
  \begin{tabular}{|l|c|}
  \hline
  \textbf{Classifier} & \textbf{Accuracy (\%)}\\
  \hline
  SVM - Linear kernel & 70.0\\
  SVM - Radial basis kernel & 72.5\\
  SVM - 1$^{st}$ order polynomial kernel &68.1\\
  SVM - 2$^{nd}$ order polynomial kernel &70.0\\
  SVM - 3$^{rd}$ order polynomial kernel &71.9\\
  SVM - 4$^{th}$ order polynomial kernel &69.4\\
  SVM - 5$^{th}$ order polynomial kernel &67.5\\
  \hline
  Weak Learners & 88.1\\
  Weak Learners \& Correlation Analysis & 85.0\\
  \hline    
  \end{tabular}
  \label{tbl:accuracy}
\end{table}
}

\newcommand{\tblmisclassification}{
\begin{table}
  \centering
  \caption{The sources of misclassification for the emotions that achieved a TRP lower than 0.9} 
  \begin{tabular}{|l|l|c|}
  \hline
  \textbf{Emotion (TPR)} & \textbf{Misclassification Source} & \textbf{MSCR}\\
  \hline
  Romantic Love (0.70)  & Anger    & 0.10\\
                 & Joy             & 0.10\\
                 & Hate            & 0.05\\
                 & Platonic love   & 0.05\\
  \hline
  Joy (0.75)     & Romantic love     & 0.15\\
                 & Grief             & 0.05\\
                 & Platonic love     & 0.05\\
  \hline
  Hate (0.75)    & Baseline (N.E.)   & 0.15\\
                 & Reverence         & 0.10\\
  \hline  
  Platonic love (0.85)  & Grief      & 0.10\\
                        & Joy        & 0.05\\
  \hline  
  Reverence (0.85)      & Hate              & 0.10\\
                        & Baseline (N.E.)   & 0.05\\
  \hline    
  \end{tabular}
  \label{tbl:misclassification}
\end{table}
}

\begin{document}
%

\maketitle
\thispagestyle{empty}
\pagestyle{empty}

\begin{abstract}
There is an increasing consensus among researchers that making a computer emotionally intelligent with the ability to decode human affective states would allow a more meaningful and natural way of human-computer interactions (HCIs). One unobtrusive and non-invasive way of recognizing human affective states entails the exploration of how physiological signals vary under different emotional experiences. In particular, this paper explores the correlation between autonomically-mediated changes in multimodal body signals and discrete emotional states. In order to fully exploit the information in each modality, we have provided an innovative classification approach for three specific physiological signals including Electromyogram (EMG), Blood Volume Pressure (BVP) and Galvanic Skin Response (GSR). These signals are analyzed as inputs to an emotion recognition paradigm based on fusion of a series of weak learners. Our proposed classification approach showed 88.1\% recognition accuracy, which outperformed the conventional Support Vector Machine (SVM) classifier with 17\% accuracy improvement. Furthermore, in order to avoid information redundancy and the resultant over-fitting, a feature reduction method is proposed based on a correlation analysis to optimize the number of features required for training and validating each weak learner. Results showed that despite the feature space dimensionality reduction from 27 to 18 features, our methodology preserved the recognition accuracy of about 85.0\%. This reduction in complexity will get us one step closer towards embedding this human emotion encoder in the wireless and wearable HCI platforms. 
\end{abstract}

\begin{keywords}
 Correlation analysis, emotional experience, feature reduction, fusion algorithm, physiological signals, weak learners.
\end{keywords}


\section{Introduction}
Providing computers with emotional understanding along with their current mathematical-logical capabilities is considered a breakthrough in creating more intelligent and less exacerbating behaviors in Human-Computer Interaction (HCI) applications \cite{picard2004toward}. An example of an intelligent HCI is exploiting "feeling computers" in enhancing distance-education experience. In \cite{vail2014predicting}, a facial recognition software has been introduced to detect specific feelings of students such as frustration and boredom during training sessions. Among the difficult challenges in these platforms is the ambiguity in recognizing emotions only from using the taxonomy of facial behaviors. This ambiguity is mainly due to the unrecognizable facial deformations such as wrinkling of the forehead or creasing around the mouth, eyes, or nose. In addition, facial expressions can be easily manipulated by the user, which produce faked affect signs. 

Unlike the facial expressions, physiological signals can not be manipulated. This characteristic makes them a robust alternative for emotion recognition systems. Knowledge of the natural processes that occur at different scales inside our body can be obtained by exploring different physiological signals and by drawing conclusions about how these biological processes are triggered, executed and connected between each other. In turn, this physiological information can be translated to advance the design and development of assistive and augmentative technologies which are human-centric and can empower people with different social, intellectual or emotional skills \cite{ref1}. 

In particular, the aim of this paper is to explore the correlation between physiological changes and emotion experiences in order to develop a robust emotion recognition algorithm. As authors concluded in \cite{ref2}, much work remains before emotion interpretation by machine intelligence can occur at the level of human abilities. When it comes to the implementation of the emotional understanding and emotional perception in machines with high-constrained computational resources, a simple but efficient knowledge of the key features that trigger and characterize human emotions could be crucial. This knowledge would provide a framework of reference for the development of future applications in which small wearable electronics can be programmed with simple and concrete definitions about how an emotion is expected to be decoded from non-invasively accessible bodily signals.

\subsection{Related Works}
Automatic emotion recognition is a field that has gained a lot of attention in the past few decades \cite{giakomis2011autoEmotionRecog, zacharatos2014autoEmotionRecog, mower2011autoEmotionRecog}. Much of the work in this area are through the exploration of diverse patterns drawn from physiological signals, and many of these signals and features are being extracted to train and test several supervised and unsupervised emotion classification methods \cite{meuleman2013emotionAndPhy, jang20142Dmodel, wen2014emotionAndPhy}. Recent studies show that autonomic affective regulation in two direction of arousal and valance is indexed by these bodily signals such as skin conductance, respiration rate, and cardiac variables, which can be measured using standard psychophysiological methods \cite{cacioppo1990inferring}. In addition, there is good evidence that physiological activity associated with psychology or mental states can be distinguished and systematically organized \cite{bradley2000emotion}. For example, electro-cardiovascular (ECG), blood volume pressure (BVP), and electromyogram (EMG) activities have been used to examine the dimension of pleasure, or valence (i.e, positive and negative affect) of human subjects \cite{papillo1990principles, cacioppo2003social}. Galvanic skin response (GSR) activity has been also shown to be associated with task engagement \cite{pecchinenda1996affective}. 

Although human physiological response to emotion has been subject of research for several years, a deeper understanding is needed to completely describe the relation between human emotional experience and each source of biosignals \cite{quigley2014there}. In addition, many efforts have also been focused on distinguishing and classifying human emotions across a 2-dimensional theoretical well-known model within the field of physiology 
\cite{nardelli20152Dmodel, jang20142Dmodel, conjeti20122Dmodel}; however, rather than focusing effort towards that direction, if the classifier is robust enough to detect a selected variation of discrete emotions, the differentiation across any theoretical dimension could be also achieved as an implicit task.

On the other hand, extracting as many features as possible and focusing the effort to improve classification accuracy at any cost, might not be the optimal approach in all cases--for instance, when attempting to implement the designed classifier in a wearable platform with very limited computing/power resources for a specific HCI application \cite{picard2001toward}. Therefore, another interesting direction to explore within the automatic emotion recognition paradigm is to enhance the optimal feature selection, feature reduction, or feature transformation methods in order to cost efficiently (in terms of power, speed, storage, etc.) exploit the related information content of human physiological signals.

\subsection{Our Contribution}
The present work extracts proper attributes from EMG, BVP, and GSR signals and feeds them to an innovative fusion-based classier to decode the correlation between the emotional experiences in different affective categories and these physiological signal modalities. \figref{signals} shows these three signals during eight discrete emotional experiences recorded over 20 minutes, which are obtained from \cite{ref1}. Our automatic emotion decoding approach is based on  output fusion of three weak learners, each built upon features extracted from the specific physiological signal modality, EMG, BVP, or GSR. The fusion algorithm is implemented to consolidate the prediction weights obtained from each of the weak learner, which are classifiers based on the linear  discriminant analysis (LDA). Furthermore, a feature selection approach based on correlation analysis is performed to reduce the complexity of the algorithm implementation while keeping the classification performance in an acceptable range.

This paper contribution capitalizes the effect of a fusion algorithm which extracts highly relevant attributes from each modality and fuse the prediction outputs rather than mixing all of the data at the same time into a single classifier. 
In addition, our proposed emotion decoding system provides the benefit of lower computational complexity by simplifying each linear weak learner through feature space dimensionality reduction by discarding highly correlated features.

\figsignals

\section{Methodology}
The general scheme for evaluating our alternative suggestion for emotion recognition from diverse physiological signals is exposed in \figref{weakAlg}. Given 19-day observations (training set), the objective is to recognize the corresponding emotions from the remaining 1-day observation (test set). A 20-fold with leave one fold out cross validation algorithm was performed in order to reduce the observation-dependent predication error. Also, the current implementation was done using MATLAB R2015b.

\figweakAlg
\subsection{Database Description}
The database provided in \cite{ref1} for research purposes, contains recordings of four physiological signals: Electromyogram (EMG), Blood Volume Pressure (BVP), Galvanic Skin Response (GSR), and Respiration (RSP) during 8 emotional states: (1) Baseline-No emotion (N.E.), (2) Anger, (3) Hate, (4) Grief, (5) Platonic Love, (6) Romantic Love, (7) Joy, and (8) Reverence. We will use these numbers as the emotion's IDs throughout this paper.

Body signal recordings were collected at a sampling rate of 20Hz for a 25-minute time period during 30 days (one observation recorded per day) \cite{ref2}. The following sensors were used for these recording: (1) a triode electromyogram measuring facial muscle tension along the masseter; (2) a photoplethysmyograph measuring blood volume pressure placed on the tip of the ring finger of the left hand; (3) a skin conductance sensor measuring electrodermal activity from the middle of the three segments of the index and middle fingers on the palm-side of the left hand; and (4) a Hall effect respiration sensor placed around the diaphragm.

The data was recorded from a female healthy graduate student with two years of acting experience and training in visualization. The participant sat in a quiet workspace early each day, at approximately the same time of day, and tried to experience eight affective states with the aid of a \textit{Sentograph}, which is a computer controlled prompting system based on the protocol for eliciting emotion developed by Clynes \cite{ref2, cacioppo2000}. According to \cite{ref2, cacioppo2000, hama1990}, the Clynes protocol has three features that contribute to help the participant feel the emotions and make the scenario appropriate for physiological data collection: \textit{(i)} it sequences eight emotions in a way that makes the transition from one emotion to another easier; \textit{(ii)} it engages physical expression by asking the participant to push a finger against a button with a dual axis pressure sensor in an expressive way that also limits the introduction of motion artifacts; \textit{(iii)} it prompts the participant to repeatedly express the same emotion during an approximately three minute interval, at a rate dependent on the emotion in order to intensify the emotional experience.

Approximately, in a third of the 30 days for which the data was collected, either one or more sensors failed during some portions of the 25-minute experiment. Therefore, two overlapping datasets (I and II) were constructed from the complete or nearly-complete sessions.

Specifically, for the scope of the present work, we are using dataset II with 3 out of the 4 physiological signals (which are EMG, BVP, and GSR), with all the 8 emotional states previously mentioned. Dataset II is the larger dataset, comprised of 20 days in which these three sensors did not fail during any part of the experiment, and which according to \cite{ref1} an average of 10\% gain in emotion recognition performance was reported when compared to the dataset I. For all of these 20 days, all of the samples available for each emotion were used plus additional transitional regions to avoid the bias and to maximize the available data for training and validation purposes.

\subsection{Database Challenge}
The structure of this dataset could picture a real scenario in which this work is intended to perform: an accurate emotion recognition when provided with a single observation to test the classifier accuracy. However, although the dataset II is comprised of a 20-day data collection scenario and each day represents approximately a 25-minute recording, the single observation allowed for testing turns this classification task very prone to observation-dependent errors. Therefore, we applied the leave-one-out cross validation scheme by averaging the prediction errors to correct for these errors.

\subsection{Signal Pre-processing: Filtering and Smoothing}
\figref{signals} presents a graphical example of the raw and filtered signals for each physiological input given the eight emotional states under study. The description of the filtering and smoothing processes applied to each of these signals is included below:
\begin{itemize}
\item EMG: This signal was filtered using a first order low-pass Butterworth filter with cutoff frequency of 10 Hz. The smoothed version of the signal was obtained by computing the average of the upper and lower envelope of the filtered signal. A detailed explanation that justifies the pertinence of using a low sampling frequency to record EMG signals is provided in \cite{ref2}.
\item BVP: The signal was filtered using a first order low-pass Butterworth filter with cutoff frequency of 19 Hz. No envelope smoothing process was applied to avoid loss of relevant information. 
\item  GSR: The signal was filtered using a first order low-pass Butterworth filter with cutoff frequency of 19 Hz (due to the fact that GSR signals are expected to be observed at 20 Hz). After filtering, the signal was smoothed by computing the average of the upper and lower envelope of the filtered signal.
\end{itemize}

\figfusionAlg

In order to compensate the nonlinear phase distortion introduced by the Butterworth filter, specially around the cut-off frequencies, the coefficients obtained from the original Butterworth filter were applied to the signal using a zero-phase digital filter known as \textit{filtfilt} in MATLAB.
\textit{filtfilt} performs filtering by processing the input data in both the forward and reverse directions: after filtering the data in the forward direction, it reverses the filtered sequence and runs it back through the filter \cite{filtfiltMatlab}. By this mechanism, the output signal achieves the desired zero-phase behavior. However, as stated in \cite{filtfiltMatlab}, one of the disadvantages to be noted is that \textit{filtfilt} would double the order of the original Butterworth filter.

Finally, a Min-Max scaling procedure was performed across the 20-day measurements in order to avoid the effects of human initial statics in the resulting feature space. This step is highly recommended for the robust performance of the classifier.

\subsection{Feature Extraction}
According to a literature review performed from \cite{ref2, ref3, ref4, ref5, ref6, ref7}, nine features were extracted for each physiological signal including time, frequency, statistical and spectral relevant characteristics including: (1) max value, (2) min value, (3) number of peaks, (4) mean: first statistical moment, (5) variance: second statistical moment, (6) kurtosis: forth statistical moment, (7) entropy, (8) signal power, and (9) signal spectral power. 

\subsection{Classifier Design}
The blocks included on the right side of \figref{weakAlg} explain the classifier design steps. Here, the proposed approach is based on specialized weak learners for each of the three physiological signals under study (EMG, BVP, GSR). A linear discriminant analysis (LDA) classifier was implemented as the weak learner for each signal. LDA is simple to calculate from data, which implies less complexity and also is reasonably robust, even when the classes do not behave as normal distributions \cite{kuncheva2004ldc, duda2001ldc}. The aim of a linear discriminant classifier is to find decision rules $g_i(x)$ in terms of the minimum total error of classification and a monotonic transformation of the posterior probabilities $P(c_i|x)$:

\begin{equation} 
\label{eqn:lda_1}
g_i(x) = \ln{P(c_i|x)} \;\;\;  i=1, ... , 8. 
\end{equation}
where each of the eight emotions are considered as the target class $c_i$, and $x$ is the set of given observation. Let's assume that each class has multivariate normal distribution and all classes have the equal covariance matrix, $\Sigma$, but distinct mean values, $\mu_i$. By Bayes theorem, $g_i(x)$, the joint posterior probability for 8 emotions can be written as a linear system:

\begin{align} 
\label{eqn:lda_2}
  g_i(x)=W_{io}+W_{i}^{T}x &\\
\text{where:} & \nonumber\\
&W_{i}= {\Sigma}^{-1}{\mu_i}  \nonumber \\
& W_{io} = -\frac{1}{2}{\mu_i}^{T}{\Sigma}^{-1}{\mu_i} + \ln {P(c_i)}   \nonumber 
\end{align}
and in a single LDA classifier, $x$ belongs to emotion class $c_i$ if $g_i(x)>g_j(x), \forall i \neq j$.

However, in our classification design, rather than concluding the emotion class $c_i$, from each LDA-based weak learner, the output from each of the three weak learners, $g_i$s, is expressed as a weighted prediction vector of the 8 emotions  that  a certain  physiological  input  is  likely  to  be  constituted  of. In turn, the three weighted predictions obtained from the three weak learners are consolidated in a robust decision by means of a fusion algorithm. 

The block diagram of the proposed fusion algorithm is presented in \figref{fusionAlg}. Here, the fusion algorithm combines the individual weighted predictions from the weak learners in a likelihood weight prediction matrix, from which a mean weight prediction vector is obtained by averaging the three weights given for each emotion. In our specific case, the dimension of the likelihood weight prediction matrix is 3$\times$8 due to the 3 physiological signals under study and the 8 discrete emotions as classification outputs.
Once the likelihood weight prediction matrix is built, in order to consolidate all individual decisions from the weak learners into a robust decision, the mean value across the columns of this matrix is computed to obtain a mean weight vector. From this mean weight vector, a discrete emotion classification is obtained by selecting the emotion for which the maximum mean weight was found. Within this concept, the idea of prediction weights can be associated with the probability of decoding eight discrete emotions from a given multimodal physiological input.

\tblaccuracy

\section{Results}
In order to demonstrate the performance of the proposed classification approach, we compared its output recognition accuracy with a traditional support vector machine (SVM) classifier implemented through different kernel functions. \tblref{accuracy} shows the recognition results achieved by the traditional SVM classifiers, the proposed weak learners algorithm, and its corresponding variant to reduce the feature space by implementing a correlation analysis.
It is important also to note that these accuracy results were obtained while applying leave-one-out cross validation technique to observe a reliable average of the classification accuracy by setting each of the 20 observations as the sample test (one at a time), while the remaining 19 observations were considered as the training set for each case.

\subsection{SVM Classifiers}
As the first attempt towards decoding emotions from physiological signals, different SVM classifiers with linear, radial basis and polynomial kernel functions were implemented. The training and test sets were built up using all the extracted features from the three physiological signals. Within this traditional classifier approach where all data were mixed in a single classifier; it can be verified from \tblref{accuracy} that the best accuracy for the traditional SVM classifier was obtained through a radial basis kernel function with 72.5\% recognition accuracy. Therefore, for the following sections and result discussion, the SVM classifier with radial basis function kernel will be the one considered to represent the SVM group behavior.

\subsection{Weak Learner-Based Classifier}
Our weak learner-based classification approach was tested under the same conditions as the ones described for SVM which also involved the leave-one-out cross validation scheme. As a result, an overall recognition accuracy of 88.1\% was achieved as it can be corroborated in \tblref{accuracy}.

\subsection{Feature Selection based on Correlation Analysis}
 In order to perform feature selection without increasing the computational complexity as when translating the feature space to other dimensions (\textit{e.g.}, principle component analysis (PCA)), a feature correlation analysis was performed prior passing the features to each weak learner. In our setting, feature correlation analysis computes the correlation between each pair of features in the whole set, and allows to drop those that are correlated with a factor greater than a predefined threshold. 

In our particular case, this threshold was set to 0.8, and from a feature space of 27 features per observation per emotion, the reduced feature space resulted in 18 uncorrelated features. Note that 27 features stand for the 9 features per each of the three physiological signals under study, and all those 27 features need to be in turn computed for each of the 20 day observations during the 8 emotion segments. Under this scenario, feature reduction from 27 to 18 features per observation per emotion represents a complexity reduction.

Within this context, by using the reduced feature space for building the training and test sets for each weak learner, the overall accuracy was preserved to 85.0\%. Note that neither PCA nor LDA provide information regarding what exact features can be removed from the feature space to reduce implementation complexity. These methods transform data to a new reduced dimensional space by computing a linear weighted combination that still considers the participation of all of the original features.
\figtprFpr

\figmisclassification
\tblmisclassification
\subsection{Emotion Decoding}

This subsection focuses on presenting emotion classification results towards the concept of decoding emotional states from the physiological signals as inputs. \figref{tprFpr} shows a variant of the receiver operating characteristic (ROC) curve, which is specialized to demonstrate true positive rate (TPR) and false positive rate (FPR) of a multi-class classifier. TPR represents the successful classification rate of a given class (emotion), and FPR represents the corresponding misclassification rate of that class among others. From the obtained curves in \figref{tprFpr}, it is clear that the proposed weak learner-based classifier with and without correlation analysis classifier achieve a higher TPR and lower FPR when compared to the SVM (with radial basis kernel function) approach.

\figref{tprFpr} indicates that the following emotion categories score a TPR lower than 0.9: romantic love, joy, hate, platonic love, and reverence. Based on this outcome, we explored which emotional sources provoked the TPR to fall below 0.9.

\figref{misclassification} shows a detailed overview about the sources which provoked misclassification for a given target emotion. Each subplot in \figref{misclassification} displays the misclassification rate (MSCR) for each of the eight target emotions under analysis when compared with the other seven remaining emotions (differentiated by their IDs). Specifically in our context, for a given target emotion, MSCR means how many times each of the other emotions provoked the classifier to misclassify the target emotion as one of them. Once again, we can corroborate that our proposed algorithm outperforms SVM, for which the MSCR is always higher.

\tblref{misclassification} also provides a key summary of the emotional sources of missclassification for each of the five emotions with TPR lower than 0.9. The entries of the table are organized from lower to higher TPR and from higher to lower MSCR.

\subsection{Discussion}

\figref{misclassification} and \tblref{misclassification} imply that the current features computed for the emotions with TPR lower than 0.9 ( \textit{romantic love, joy, hate, platonic love, reverence}) are not sufficient enough to provide a meaningful distinction among these classes; and therefore, more relevant features should be computed. In turn, when the work switches towards finding relevant features to improve the distinguishability among specific classes, a misclassification analysis like the one suggested in \figref{misclassification} might be useful to evaluate if the new set of given features are the correct ones.  

Moreover, the benefit of providing an alternative visualization method for detecting the sources of misclassification for each emotion is of high usability when addressing the task of re-computing the feature set for a given class in order to improve individual class recognition accuracy.

One interesting question that arises from the physiological dataset under study, is why the LDA classifier outperforms the SVM approach? - Even when considering the case of SVM with a linear kernel function, which did not achieved the best behaviour among the different kernels tested for SVM.
Since the data shows to be linearly separable, the SVM approach of mapping data into a higher dimensional space in which it would be linearly separable turns out to be redundant. If the data was already observed to be linearly separable, LDA represents a good approach towards classification without overfitting the classification model \cite{lin2011svm}. The strength of our model then is mainly due to the fusion of  multiple decisions provided by different weak learners each tuned for a specific input modality.



\section{Conclusion and Future Work}
The proposed classification approach based on specialized weak learners for each physiological signal reported a higher accuracy than the common approach of designing a unique classifier intended to learn features from all the input signals. The suggested approach also highlights modularity benefits when considering to add extra information, without increasing the complexity granted in the case of employing a single classifier.

In fact, the proposed fusion algorithm is an alternative boosting approach for consolidating multiple decisions provided by different weak learners in a strong classification output. By working with prediction scores as outputs of each weak learner, we are considering more than one discrete emotion that a single test input is likely to be constituted of; and this flexibility obtained from the prediction scores can be extremely relevant when a final classification decision is made by combining the criteria provided by diverse weak learners. 

Furthermore, simple methods such as feature correlation analysis can be very helpful and powerful tools at the same time when the objective is to analyze redundancy across the feature set. By performing this prior correlation study, redundant features can be pruned, meaning that a fine-grained feature selection can be passed as input to each of the weak learners to improve individual classification accuracy. Feature pruning referred as feature selection is of great interest when simple learning relations among features need to be drawn for classifier implementations where the computational resources are very limited.

It is important to also note that the proposed feature reduction approach based on correlation analysis automatically handles the procedure of removing features. In other words, due to the fact that for each training set the suitable features might be different, the number of features that are selected for each set changes accordingly.

Finally, a great amount of further steps are required in order to fully exploit the capabilities of the proposed system. For instance, one interesting question regarding feature selection within the correlation analysis is to include a penalization not just for redundancy of information, but for considering the case of information confusion or mismatch between features. Another interesting direction is to explore the coherence of valence and arousal dimensions in which emotions are generally explained by a theoretical model with respect to the optimal number of orthogonal dimensions that techniques such as PCA may suggest to preserve a desired amount of information within an specific variance of the data.

\balance
\bibliographystyle{IEEEtran}
\small
\bibliography{paper}

\end{document}